\documentclass[twocolumn]{jpsj2}

\def\runauthor{Youichi {\sc Yanase} and Masao {\sc Ogata}}

\title{ 
Kinetic Energy, Condensation Energy, Optical Sum Rule \\ 
and Pairing Mechanism in High-T$_{\rm c}$ Cuprates
} 

\author{Youichi {\sc Yanase}\footnote{E-mail:
yanase@hosi.phys.s.u-tokyo.ac.jp} and Masao {\sc Ogata}}

\inst{Department of Physics, University of Tokyo, Tokyo 113-0033}

\recdate{Today 2004}

\abst
{
 The mechanism of high-$T_{\rm c}$ superconductivity is investigated 
with interests on the microscopic aspects of the condensation energy. 
 The theoretical analysis is performed on the basis of the 
FLEX approximation which is a microscopic description of 
the spin-fluctuation-induced-superconductivity. 
 Most of phase transitions in strongly correlated electron system arise 
from the correlation energy which is copmetitive to the kinetic energy. 
 However, we show that the kinetic energy cooperatively induces 
the superconductivity in the underdoped region.  
 This unusual decrease of kinetic energy below $T_{\rm c}$ is induced 
by the feedback effect. The feedback effect induces the magnetic 
resonance mode as well as the kink in the electronic dispersion, and 
alters the properties of quasi-particles, such as mass renormalization 
and lifetime. 
 The crossover from BCS behavior to this unusual behavior occurs 
for hole dopings. 
 On the other hand, the decrease of kinetic energy below $T_{\rm c}$ 
does not occur in the electron-doped region. 
 We discuss the relation to the recent obserbation of the violation of 
optical sum rule. 
}

\kword
{
High-T$_{\rm c}$ superconductivity; spin fluctuation; 
feedback effect; kinetic energy; condensation energy; optical sum rule
}

\begin{document}
\sloppy
\maketitle

\newcommand{\eli}{$\acute{{\rm E}}$liashberg }
\renewcommand{\k}{\mbox{\boldmath$k$}}
\newcommand{\q}{\mbox{\boldmath$q$}}
\newcommand{\Q}{\mbox{\boldmath$Q$}}
\newcommand{\kk}{\mbox{\boldmath$k'$}}
\newcommand{\e}{\varepsilon}
\newcommand{\ee}{\varepsilon^{'}}
\newcommand{\s}{{\mit{\it \Sigma}}}
\newcommand{\J}{\mbox{\boldmath$J$}}
\newcommand{\vv}{\mbox{\boldmath$v$}}
\newcommand{\Jh}{J_{{\rm H}}}
\newcommand{\LL}{\mbox{\boldmath$L$}}
\renewcommand{\SS}{\mbox{\boldmath$S$}}
\newcommand{\Tc}{$T_{\rm c}$ }
\newcommand{\Tcf}{$T_{\rm c}$}
\newcommand{\Co}{${\rm Na_{x}Co_{}O_{2}} \cdot y{\rm H}_{2}{\rm O}$ }
\newcommand{\Cof}{${\rm Na_{x}Co_{}O_{2}} \cdot y{\rm H}_{2}{\rm O}$}
\newcommand{\tgf}{$t_{\rm 2g}$-orbitals}
\newcommand{\tg}{$t_{\rm 2g}$-orbitals }
\newcommand{\av}{\mbox{\boldmath${\rm a}$} }
\newcommand{\bv}{\mbox{\boldmath${\rm b}$} }
\newcommand{\avf}{\mbox{\boldmath${\rm a}$}}
\newcommand{\bvf}{\mbox{\boldmath${\rm b}$}}
\newcommand{\egf}{$e_{\rm g}$-Fermi surface }
\newcommand{\egff}{$e_{\rm g}$-Fermi surface}
\newcommand{\agf}{$a_{\rm 1g}$-Fermi surface }
\newcommand{\agff}{$a_{\rm 1g}$-Fermi surface}

\section{Introduction}

 The mechanism of high-\Tc superconductivity in cuprate 
materials~\cite{rf:bednortz} has been one of the most 
appealing subject in the condensed matter physics 
over the last couple of decades. 
 Through intensive studies from theoretical and experimental 
researchers, the ``magnetic mechanism'' is believed most predominantly.

 The ``magnetic mechanism'' is represented by the spin fluctuation 
theory which takes into account the interaction between 
quasi-particles exchanging the spin fluctuations.~\cite{
rf:moriya1990,rf:monthoux1991} 
 Among the microscopic descriptions beyond the original 
phenomenology,~\cite{rf:moriya1990,rf:monthoux1991,rf:moriyaAD,
rf:chubukovreview} the fluctuation-exchange (FLEX) 
approximation is adopted most widely.~\cite{rf:FLEX,rf:monthouxFLEX,
rf:paoFLEX,rf:dahmFLEX,rf:langerFLEX,rf:koikegamiFLEX,
rf:takimotoFLEX,rf:takimotoFLEX2} 
 The qualitative validity of this theory is highly expected from 
weak to intermediate coupling region.~\cite{rf:yanasereview} 
 For example, the spin fluctuation theory is robust for the vertex 
corrections arising from the non-RPA terms~\cite{rf:hotta,rf:nomura} 
as well as those from the multiple spin fluctuation exchange 
terms.~\cite{rf:monthoux1997,rf:chubukov1997} 
 On the other hand, in the expansion from the strong coupling limit, 
the resonating valence bond (RVB) 
theory~\cite{rf:andersonbook,rf:fukuyamaRVBreview,rf:nagaosa} 
and some numerical methods~\cite{rf:yokoyama,rf:gros,rf:dagotto,
rf:dagottoreview,rf:sorella} 
have concluded the $d_{x^{2}-y^{2}}$-wave superconductivity where the 
super-exchange interaction plays an essential role.

 In the spin fluctuation theory, \eli equation is used for an analysis 
of the superconductivity. 
 The \eli equation provides a clear understanding for the 
mechanism of superconductivity on the basis of the BCS picture. 
 Then, the attractive interaction leading to the Cooper pairing is 
represented by the irreducible four point vertex. 
 The momentum dependence of this vertex induces the non-$s$-wave 
superconductivity. 
 In the spin fluctuation theory for high-\Tc cuprates, the irreducible 
four point vertex is described by the anti-ferromagnetic 
spin fluctuation and the attractive interaction is most effective 
in the $d$-wave channel.

 The goal of this paper is to investigate the mechanism of 
high-\Tc superconductivity from another point of view. 
 We study {\it how the energy is gained below \Tc owing to the 
superconductivity}. 
 Of course, the ground state is determined as a result of 
the energetic optimization. 
 The understanding from the energetics will be complementary to 
the analysis of the interaction leading to the pairing. 
 Some interesting aspects are clarified from this point of view.

 This study is partly motivated by the theoretical proposal for 
``kinetic energy driven pairing''~\cite{rf:andersonbook,rf:hirsh1992,
rf:imada,rf:hirsh2002,rf:chakrabarty} 
and by the recent experimental supports for this 
proposal.~\cite{rf:molegraaf,rf:santander} 
 In the conventional BCS theory, the kinetic energy increases owing 
to the superconductivity. 
 This increase is slightly over-compensated by the decrease of 
correlation energy. 
 Contrary to the BCS theory, the ``kinetic energy driven pairing'' 
attributes the mechanism of superconductivity to the gain of 
kinetic energy. 
 This mechanism has been considered to be highly unconventional and 
the discrepancy to the spin-fluctuation-induced-superconductivity 
has been noted.~\cite{rf:imada,rf:hirsh2002,rf:chakrabarty} 
 On the other hand, the consistency to the RVB 
theory has been discussed, where the superconducting transition 
is triggered by the coherence of charge 
carriers.~\cite{rf:andersonbook,rf:anderson2000,rf:lee1999} 
 As for numerical studies, the dynamical cluster 
approximation~\cite{rf:maier} and variational Monte Carlo 
simulation~\cite{rf:yokoyamakinetic} for Hubbard model have shown 
a decrease of kinetic energy owing to the superconductivity, and then 
implications for the RVB state have been noted.

 In this paper, we study these problems on the basis of 
the microscopic and strong coupling theory on 
the spin-fluctuation-induced-superconductivity, 
namely the FLEX approximation. 
 This subject has been investigated by several 
phenomenological theories assuming the non-Fermi liquid normal 
state,~\cite{rf:norman2000,rf:norman2002} 
spin-Fermion coupling,~\cite{rf:haslingerrapid,rf:haslingerfull} 
superconducting phase fluctuation,~\cite{rf:eckl} 
and electron-phonon coupling.~\cite{rf:knigavko} 
 In contrast to these theories, the FLEX approximation is a 
``conserving approximation'' which is highly suitable for a discussion  
of thermodynamic properties. 
 In the conserving approximation formulated by Luttinger and 
Ward,~\cite{rf:luttinger,rf:baym} 
all quantities are self-consistently derived from the differential 
of thermodynamic potential without any phenomenological assumption. 
 Unphysical results inherent in the phenomenological theory 
are considerably excluded in the microscopic theory adopted here. 
 It should be stressed that highly careful treatment is needed for 
thermodynamic properties rather than for magnetic or 
single-particle properties.  
 This is partly because the condensation energy of superconductivity 
is much smaller than the energy scale of electrons. 
 For example, the condensation energy is in the order of $0.1$meV, 
while the band width is in the order of $1$eV.

 Note that kinetic energy along {\it c}-axis has attracted 
much interests in the early stage because it is related to the 
'interlayer tunneling mechanism' (ILT) proposed by Anderson.~\cite{rf:ILT} 
 Optical measurements have supported the decrease of {\it c}-axis 
kinetic energy.~\cite{rf:basov1999,rf:katz,rf:basov2001} 
 However, it has been shown that the gain of {\it c}-axis kinetic 
energy is much smaller than the condensation energy,~\cite{rf:kirtley,
rf:moler,rf:vandermarel} and therefore this subject is not essential 
for the mechanism of superconductivity. 
 In this paper we focus our attention on the kinetic energy 
along the plane.

 In \S2, we formulate a conserving approximation in the superconducting 
state and provide the expressions of FLEX approximation. 
 Results on the kinetic energy are shown in \S3.1. 
 We show that the kinetic energy decreases below \Tc in the 
under-doped region while it increases like BCS theory in the 
over-doped region and in the electron-doped region. 
 It will be stressed that the concepts of 
``spin-fluctuation-induced-superconductivity'' 
and ``kinetic energy driven pairing'' are not incompatible. 
 The relation between the kinetic energy and the optical sum rule 
is discussed in \S3.2. 
 In \S3.3, we discuss thermodynamic properties in more details. 
 Then, we propose another interpretation of the condensation energy 
by considering the free energy arising from the spin fluctuation. 
 Some discussions are given in \S4.

\section{Thermodynamic Property and FLEX Approximation below \Tc}

 In this paper, we analyze the two-dimensional Hubbard model 
which is expressed as,  
\begin{eqnarray}
  \label{eq:Hubbard-model}
  && \hspace{-5mm} H=\sum_{{\k},\sigma} \varepsilon(\k) 
  c_{{\k}\sigma}^{\dag}c_{{\k}\sigma}
  + U \sum_{i} n_{{i}\uparrow} n_{{i}\downarrow}. 
\end{eqnarray}
 We consider the square lattice and choose the following tight-binding 
dispersion, 
\begin{eqnarray}
  \label{eq:high-tc-dispersion}
  &&  \hspace{-5mm} \varepsilon(\k)=-2t(\cos k_{\rm x}+\cos k_{\rm y})
  +4t'\cos k_{\rm x} \cos k_{\rm y}. 
\end{eqnarray}
 In the following, the unit of energy is chosen as $2t=1$. 
 The next nearest neighbor hopping $t'$ is necessary and sufficient to 
reproduce the Fermi surface of high-\Tc cuprates. 
 The typical value is estimated to be $t'/t=0.1 \sim 0.4$.  
 Qualitative results in this paper are not altered by this value. 
 We fix $t'/t=0.25$ in the hole-doped region and $t'/t=0.35$ 
in the electron-doped region, respectively. 
 The concentration of hole doping is expressed as $\delta=1-n$ where
$n$ is the density of electrons per sites.

 In the superconducting state, statistical quantum field theory is 
described by normal and anomalous Green functions, $G(k)$ and $F(k)$. 
 The Dyson-Gorkov equation describes the Green functions through 
the normal and anomalous self-energies, which are denoted as 
$\Sigma^{\rm n}(k)$ and $\Delta(k)$, respectively. 
\begin{eqnarray}
 && \hspace{-10mm}
\left(
 \begin{array}{cc}
   G(k) & F(k) \\
   F^{\dag}(k) & -G(-k)
 \end{array}
 \right)=
\nonumber \\
&& \hspace{-10mm}
 \left(
 \begin{array}{cc}
   G^{(0)}(k)^{-1} - \Sigma^{\rm n}(k) & \Delta(k) \\
   \Delta^{*}(k) & -G^{(0)}(-k)^{-1} + \Sigma^{\rm n}(-k)
 \end{array}
 \right)^{-1}.
\end{eqnarray}
 Here, $G^{(0)}(k)$ is the Green function in the non-interacting case, 
\begin{eqnarray}
G^{(0)}(k)=\frac{1}{{\rm i}\omega_n - \varepsilon(\k) +\mu}, 
\end{eqnarray} 
where $\mu$ is the chemical potential. 
 The superconducting gap $\tilde{\Delta}(\k)$ is obtained 
by the anomalous self-energy as $\tilde{\Delta}(\k)=z(\k) |\Delta(k)|$ 
where $z(\k)^{-1}=1-
\partial {\rm Re} \Sigma^{\rm R}(\k,\omega)/\partial \omega|_{\omega=0}$.

 In order to discuss the thermodynamic quantities, we first formulate 
a general expression for the thermodynamic potential 
in the superconducting state and derive the self-energy, momentum 
distribution function and kinetic energy on the basis of the functional 
derivatives. 
 In the following, we describe the formulation in case of 
the spin singlet pairing.

 The conserving form of the thermodynamic potential in the normal state 
was formulated by Luttinger and Ward~\cite{rf:luttinger} 
and developed by Baym and Kadanoff.~\cite{rf:baym} 
 Then, the self-energy is obtained by the functional derivative of 
generating function $\Phi$ as $\Sigma^{\rm n}_{\sigma}(k) = 
\delta \Phi[G_{\sigma}]/\delta G_{\sigma}(k)$. 
 The thermodynamic potential is obtained by the generating function and 
self-energy as, 
\begin{eqnarray} 
&&\hspace{-10mm} \Omega(T,\mu) = \Omega_{0}(T,\mu) - \sum_{\sigma}\sum_{k} 
[\log\{\frac{\Sigma^{\rm n}_{\sigma}(k) - G^{(0)}(k)^{-1}}
{- G^{(0)}(k)^{-1}} \} 
\nonumber \\
&&\hspace{5mm}
 + G_{\sigma}(k) \Sigma^{\rm n}_{\sigma}(k)] + \Phi[G_{\sigma}]. 
 \label{eq:ithermodynamic-nomal} 
\end{eqnarray} 
 Here, $\Omega_{0}(T,\mu)= -2 T \sum_{\k} log[1+\exp\{-\beta(\e(\k)-\mu)\}]$ 
is the thermodynamic potential in the non-interacting case. 
 Although we have formally introduced the index of spin $\sigma$,  
indeed, $G_{\sigma}(k)=G(k)$ and 
$\Sigma^{\rm n}_{\sigma}(k)=\Sigma^{\rm n}(k)$ 
since we consider the paramagnetic state or spin singlet 
superconducting state.

 It is straightforward to generalize this formulation 
to the superconducting state. 
 We obtain the normal and anomalous self-energies from 
the generating function $\Phi[G_{\sigma},F,F^{\dag}]$ as, 
\begin{eqnarray}
 \label{eq:derivative}
 && \hspace{-10mm}
 \Sigma^{\rm n}_{\sigma}(k) = \frac{\delta \Phi}{\delta G_{\sigma}(k)}, 
\\
 \label{eq:derivative2}
 &&\hspace{-10mm}
 \Delta(k) = -\frac{\delta \Phi}{\delta F^{\dag}(k)}, 
\\
 \label{eq:derivative3}
 &&\hspace{-10mm}
 \Delta^{*}(k) = -\frac{\delta \Phi}{\delta F(k)}.
\end{eqnarray}
 Note that eq.~(\ref{eq:derivative2}) (equivalently eq.~(\ref{eq:derivative3})) 
is a self-consistent equation determining the second order 
phase transition. The linearized version of eq.~(\ref{eq:derivative2}) 
has been used in order to determine the superconducting 
instability.~\cite{rf:yanasereview} 
 We obtain the general expression of thermodynamic potential as, 
\begin{eqnarray} 
  \label{eq:thermodynamic-super} 
&&\hspace{-10mm} \Omega(T,\mu) = \Omega_{0}(T,\mu) + \Omega_{\rm F}
+ \Omega_{\rm B}, 
\\
  \label{eq:thermodynamic-superF} 
&&\hspace{-10mm} \Omega_{\rm F} = 
- \sum_{k} 
[\log\{\frac{|\Sigma^{\rm n}_{\sigma}(k) - G^{(0)}(k)^{-1}|^{2} 
+ |\Delta(k)|^{2}}{|- G^{(0)}(k)^{-1}|^{2}}\} 
\nonumber \\
&& \hspace{-5mm}+ \sum_{\sigma} G_{\sigma}(k) \Sigma^{\rm n}_{\sigma}(k) 
- F(k) \Delta^{*}(k) - F^{\dag}(k) \Delta(k)], 
\\
  \label{eq:thermodynamic-superB} 
&& \hspace{-10mm} \Omega_{\rm B} = \Phi[G_{\sigma},F,F^{\dag}]|_{\rm st}. 
\end{eqnarray} 
 The derivation of 
eqs.~(\ref{eq:thermodynamic-super}-\ref{eq:thermodynamic-superB}) 
is summarized in Appendix. 
 According to eqs.~(\ref{eq:derivative}-\ref{eq:thermodynamic-superB}), 
the variational conditions with respect 
to the self-energy are satisfied as, 
\begin{eqnarray}
 \label{eq:variational-condition}
 &&\hspace{-10mm}  \frac{\delta \Omega}{\delta \Sigma_{\rm n}(k)}=
 \frac{\delta \Omega}{\delta \Delta(k)}=0. 
\end{eqnarray}
 Therefore, the self-energies obtained by 
eqs.~(\ref{eq:derivative}-\ref{eq:derivative3}) 
provide a stationary value of thermodynamic potential. 
 According to the thermodynamics, we obtain the number density as 
\begin{eqnarray}
 \label{eq:number-density}
&&\hspace{-10mm}  n = - \frac{\delta \Omega}{\delta \mu} 
= 2 \sum_{\k} n(\k),  
\end{eqnarray}
where the momentum distribution function is also obtained by 
the functional derivatives as, 
\begin{eqnarray}
 \label{eq:momentum-distribution2}
&& \hspace{-10mm} 
n(\k)=\frac{1}{2} \frac{\delta \Omega}{\delta \e(\k)}. 
\end{eqnarray}
 By performing the functional derivatives, 
eq.~(\ref{eq:momentum-distribution2}) is reduced to the usual definition 
of $n(\k)$, 
\begin{eqnarray}
&&\hspace{-8mm} n(\k)=
\sum_{\omega_n} G(k) {\rm e}^{{\rm i} \omega_n \delta}
\nonumber \\
 \label{eq:momentum-distribution}
&& = \sum_{\omega_n} [G(k)-G^{(0)}(k)] + f(\e(k)-\mu), 
\end{eqnarray}
where we have eliminated the ultra-violet divergence by 
subtracting the Fermi distribution function $f(\e(k)-\mu)$. 
 Finally, the kinetic energy is obtained as, 
\begin{eqnarray}
 \label{eq:kinetic-energy}
&&\hspace{-10mm} 
E_{\rm k} =  \sum_{\k} \e(\k) \frac{\delta \Omega}{\delta \e(\k)} 
= 2 \sum_{\k} \e(\k) n(\k). 
\end{eqnarray}

 When these relations are self-consistently satisfied in an approximation, 
the approximation is classified into the 
``conserving approximation''.~\cite{rf:baym} 
 The FLEX approximation is one of them. 
 In the following, we fix the number density $n$ instead of the 
chemical potential $\mu$. Therefore, the free energy 
$F(T,n)=\Omega+\mu n$ is a more convenient quantity. 

 Hereafter, the superscripts $^{\rm S}$ and $^{\rm N}$ 
denote the superconducting and normal states, respectively. 
 The former is defined by the stationary solution with finite value 
of $\Delta(k)$ and the latter is defined by the solution with 
$\Delta(k)=0$. 
 Both solutions satisfy the variational conditions 
eq.~(\ref{eq:variational-condition}), but only the 
former satisfies the stable condition below \Tcf.  
 The difference between normal and superconducting states are 
denoted as $\delta A=A^{\rm N}-A^{\rm S}$. 
 For example, the condensation energy is expressed as 
$\delta F=F^{\rm N}-F^{\rm S}$. 

 Before closing the general formulation, we note some analytical 
expressions for the parameter dependence of the free energy. 
 First, the number dependence of the free 
energy is given by the chemical potential $\partial F/\partial n =\mu$. 
 Therefore, the number dependence of the condensation energy 
is obtained as, 
\begin{eqnarray}
 \label{eq:number-derivative} 
\frac{\partial \delta F}{\partial n} = \mu^{\rm N} - \mu^{\rm S}. 
\end{eqnarray}
 Second, the $U$-dependence of the free energy is given 
by the running coupling constant formula which is expressed as follows, 
\begin{eqnarray}
&& \hspace{-10mm}
\frac{\partial F}{\partial U} = \frac{1}{2 U}
\sum_{k} [\sum_{\sigma} G_{\sigma}(k) \Sigma^{\rm n}_{\sigma}(k) 
\nonumber \\
&& \hspace{11mm}
- F(k) \Delta^{*}(k) - F^{\dag}(k) \Delta(k)].
\label{eq:U-derivative} 
\end{eqnarray}
 The $U$-dependence of the condensation energy is simply
obtained by the subtraction. 
 These expressions are convenient to understand the qualitative 
behaviors of the condensation energy (see \S3.3).

 The formulation of FLEX approximation has been given in
literatures.~\cite{rf:FLEX,rf:yanasereview} 
 The extension to the superconducting state is straightforward. 
 Indeed, some authors have investigated the properties in the 
superconducting state by using the FLEX approximation. 
 For instance, the temperature dependence of 
superconducting gap,~\cite{rf:monthouxFLEX,rf:paoFLEX} magnetic and 
single-particle properties~\cite{rf:paoFLEX,rf:dahmFLEX,rf:takimotoFLEX2} 
have been discussed. 
 In this paper, we analyze the thermodynamic quantities and 
their relation to the optical sum rule.

 The generating function $\Phi[G_{\sigma},F,F^{\dag}]$ is obtained  
in the FLEX approximation as, 
\begin{eqnarray}
&& \hspace{-15mm}
\Phi[G_{\sigma},F,F^{\dag}] = 
\nonumber \\
&&\hspace{-15mm} 
\sum_{q} 
[
\frac{3}{2} \log\{1- U \chi_{\rm s}^{0}(q)\}
+\frac{1}{2} \log\{1 + U \chi_{\rm c}^{0}(q)\}
\nonumber \\
&&\hspace{-10mm} 
+\frac{1}{4} U^{2}(\chi_{\rm s}^{0}(q)^{2}+\chi_{\rm c}^{0}(q)^{2})
+U (\frac{3}{2}\chi_{\rm s}^{0}(q) - \frac{1}{2} \chi_{\rm c}^{0}(q))
]. 
 \label{eq:Phi-FLEX}
\end{eqnarray}
Here, we have denoted the irreducible spin and charge susceptibilities 
as, 
\begin{eqnarray}
  \label{eq:irreducible-susceptibility}
  &&\hspace{-15mm}  
\chi_{\rm s,c}^{0}(q) = -\sum_{k} [G(k+q) G(k) \pm F(k+q) F(k)].  
\end{eqnarray} 
 Note that we have ignored the first order terms in the generating function 
since their roles are trivial and do not affect the following discussions.

 We obtain the self-energy from 
eqs.~(\ref{eq:derivative}-\ref{eq:derivative2}) as, 
\begin{eqnarray}
  \label{eq:self-energy}
  && \hspace{-15mm} 
\Sigma^{\rm n}(k) = \sum_{q} V_{\rm n}(q) G(k-q), 
\\
  \label{eq:self-energy2}
  && \hspace{-15mm} 
\Delta(k) = - \sum_{q} V_{\rm a}(q) F(k-q), 
\end{eqnarray}
where $V_{\rm n}(q)$ and $V_{\rm a}(q)$ are expressed as, 
\begin{eqnarray}
  \label{eq:vertex}
  && \hspace{-15mm} 
V_{\rm n} (q)=U^{2} [\frac{3}{2} \chi_{{\rm s}} (q)
  +\frac{1}{2} \chi_{{\rm c}} (q) 
  - \frac{1}{2}\{\chi_{\rm s}^{0}(q)+\chi_{\rm c}^{0}(q)\}], 
\\
  && \hspace{-15mm} 
V_{\rm a} (q)=U^{2} [\frac{3}{2} \chi_{{\rm s}} (q)
  -\frac{1}{2} \chi_{{\rm c}} (q)
  - \frac{1}{2}\{\chi_{\rm s}^{0}(q)-\chi_{\rm c}^{0}(q)\}]. 
\end{eqnarray}
 We have introduced the spin and charge susceptibilities 
obtained by the generalized RPA as,  
\begin{eqnarray}
 \label{eq:spin-susceptibility}
  && \hspace{-15mm} 
\chi_{{\rm s,c}}(q) = 
     \frac{\chi_{\rm s,c}^{0}(q)}{1 \pm U \chi_{\rm s,c}^{0}(q)}. 
\end{eqnarray} 
 The normal vertex $V_{\rm n}(q)$ and anomalous vertex $V_{\rm a}(q)$ 
are represented by the irreducible four point vertex 
in the particle-hole channel and in the particle-particle channel, 
respectively.

 The superconducting transition from the normal state is determined by 
the appearance of non-trivial solution of eq.~(\ref{eq:self-energy2}). 
 In analogy with the gap equation in the BCS theory, we often denote 
$V_{\rm a}(q)$ as effective interaction leading to the pairing. 
 Generally speaking, the momentum dependence of effective 
interaction results in the attractive interaction in a non-$s$-wave
channel.~\cite{rf:yanasereview} 
 In the FLEX approximation, the effective interaction $V_{\rm a}(q)$ is 
dominated by the spin fluctuation whose momentum dependence is 
favorable for the $d_{x^{2}-y^{2}}$-wave superconductivity. 
 This is the ordinary understanding on the spin-fluctuation-induced 
superconductivity.

 In this paper, we obtain another insights on the 
spin-fluctuation-induced superconductivity which are 
given by the analysis of energetics. 
 The free energy is described as $F=E_{\rm k}+E_{\rm cr}- T S$ where 
$E_{\rm cr}=<U n_{{i},\uparrow} n_{{i},\downarrow}>$ is the correlation 
energy and $S$ is the entropy. 
 At $T=0$, the condensation energy is obtained by the kinetic energy and 
correlation energy as $\delta F=\delta E_{\rm k} + \delta E_{\rm cr}$. 
 In \S3.1, we show that the kinetic energy increases the condensation 
energy cooperatively with the correlation energy. 
 This is in sharp contrast with the weak coupling BCS theory where 
the kinetic energy remarkably decreases the condensation energy.

 Note that the FLEX approximation provides a reasonable value of \Tc 
not only in the hole-doped region but also in the electron-doped region. 
 \Tc and doping region with superconducting order is 
very small in the electron-doped region~\cite{rf:manske,
rf:yanaseFLEXPG,rf:ekondo,rf:hirashima} owing to the small DOS and 
the localized character of spin fluctuation in the momentum space.

 It should be noticed that the FLEX approximation does not explain the 
pseudogap phenomena in the normal state of under-doped region. 
 The superconducting fluctuation should be included to explain 
the pseudogap phenomena in this framework.~\cite{rf:yanaseFLEXPG,
rf:yanaseTRPG,rf:yanasebeyond1} 
 However, it has been shown that the effects of superconducting 
fluctuation on the electronic state is rapidly suppressed 
below \Tcf.~\cite{rf:yanaseSC}
 Therefore, we believe that the FLEX approximation is appropriate 
for a description of superconducting state. 
 We note that some interesting phenomena, such as 
magnetic resonance peak~\cite{rf:takimotoFLEX2} and 
kink in the electronic dispersion~\cite{rf:yanaseunpublished}, 
are well explained within the FLEX approximation.

 We have used the notations $\sum_{k}=T/N \sum_{\omega_n,\k}$ 
and $\sum_{q}=T/N \sum_{\Omega_n,\q}$ where 
$\omega_n=(2 n + 1) \pi T$, $\Omega_n=2 n \pi T$, 
$T$ is the temperature and $N$ is the number of sites. 
 The unit $\hbar=c=k_{{\rm B}}=1$ is used through this paper. 
 Eqs.~(\ref{eq:irreducible-susceptibility}), (\ref{eq:self-energy}) and 
(\ref{eq:self-energy2}) are estimated by using the 
fast Fourier transformation (FFT).

\section{Results}

 Before showing the results, we note some cares involved in 
the numerical calculation because a careful treatment is highly 
needed for an estimation of thermodynamic quantities. 
 This is partly because the electronic states far below Fermi level 
essentially contribute to the free energy, and partly 
because the condensation energy is a very small value 
in the order of $ 10^{-4} \sim 0.1$meV.

\begin{figure}[htbp]
  \begin{center}
\includegraphics[height=12cm]{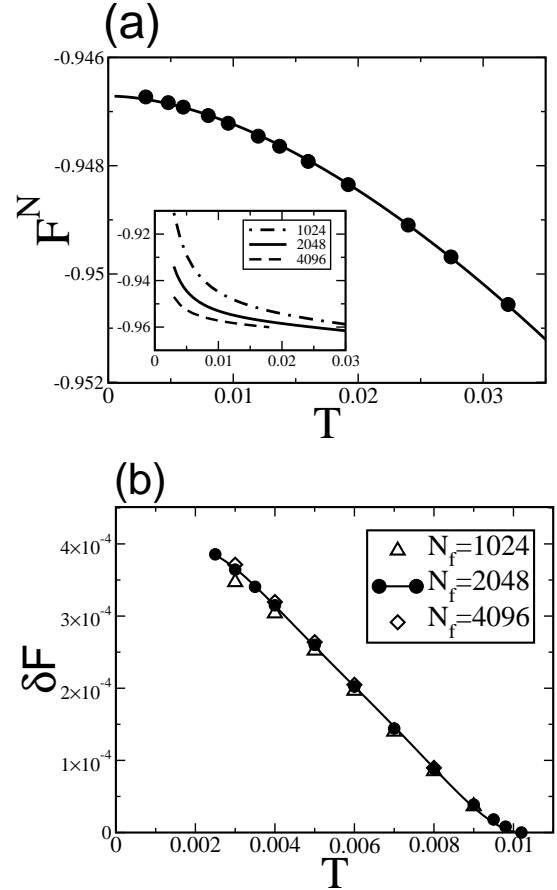}
    \caption{(a) Temperature dependence of free energy in the 
             normal state calculated by fixing the cut-off frequency. 
             The solid line is a fitting curve with use of the 
             function $F^{\rm N}=a + b T^{2}+c T^{2} \log T$. 
             The inset shows the same quantity with use of fixed 
             $N_{\rm f}=1024, 2048$ and $4096$. 
             (b) Temperature dependence of $\delta F$ 
             for various $N_{\rm f}$. 
             We choose the parameters as $\delta=0.1$ and $U/t=4.2$. 
             }
    \label{fig:cut-off}
  \end{center}
\end{figure}
 
 We divide the first Brillouin zone into $N \times N$ and 
take $N_{\rm f}$ Matsubara frequency. 
 The numerical inaccuracy mainly arises from the cut-off of 
Matsubara frequency. 
 It should be noticed that we have introduced expressions 
without ultra-violet divergence in eqs.~(\ref{eq:thermodynamic-super})
and (\ref{eq:momentum-distribution}).  
 This procedure remarkably improves the numerical accuracy. 
 However, further care is needed for a temperature dependence of 
thermodynamic quantities.  
 If we fix the number of Matsubara frequency $N_{\rm f}$, 
the cut-off of frequency depends on the temperature as 
$\omega_{\rm c}=(N_{\rm f}-1)\pi T$. 
 This induces an artificial temperature dependence which 
may smear the intrinsic temperature dependence. 
 This difficulty is very serious for the free energy, as is shown in 
Fig.~\ref{fig:cut-off}(a). 
 The main figure shows the temperature dependence of free energy 
$F^{\rm N}$ calculated with the cut-off frequency fixed to be 
$\omega_{\rm c} = 38.6 \sim 10 W$. 
 Then, the free energy is well fitted by the function 
$F^{\rm N}=a + b T^{2}+c T^{2} \log T$, 
which is consistent with the nearly anti-ferromagnetic Fermi liquid 
state.~\cite{rf:moriyaAD,rf:chubukovreview} 
 On the other hand, the free energy obtained by fixing $N_{\rm f}$ 
shows much larger temperature dependence 
(see inset in Fig.~\ref{fig:cut-off}(a)) 
indicating the violation of thermodynamic third law. 
 Thus, we have to fix the cut-off frequency instead of $N_{\rm f}$ 
in order to obtain appropriate results. 
 Unfortunately, owing to the computational constraint arising 
from the FFT, it is troublesome to fix the cut-off frequency 
$\omega_{\rm c}=(N_{\rm f}-1)\pi T$ for various temperatures.

 However, this difficulty does not matter for 
the differences between normal and superconducting states. 
 This is because the superconductivity affects the low energy states while 
high energy states are not sensitive to the superconductivity. 
 For instance, we show the temperature dependence of $\delta F$ 
in Fig.~\ref{fig:cut-off}(b). 
 It is clearly shown that $\delta F$ depends on $N_{\rm f}$ only 
slightly, and the fixed-$N_{\rm f}$ calculation is valid. 
 Note that $N_{\rm f}$-dependence of the calculated free energy is still 
in the order of $10^{-3}$ at $T=0.005$. 
 However, the difference $\delta F$ is estimated to 
an accuracy of $10^{-6}$. 
 This circumstance is in common with the momentum distribution function, 
kinetic energy and internal energy. 
 We have confirmed that $2048$ Matsubara frequency is sufficient for 
the following results. 
 We show the results obtained by using $64 \times 64$ meshes or 
$128 \times 128$ meshes in the first Brillouin zone 
in the hole-doped case. 
 We have confirmed that the finite size effects are negligible in 
these calculations. 
 In the electron-doped region, we use $256 \times 256$ meshes in order 
to ensure the numerical accuracy.

\subsection{Kinetic energy}

\begin{figure}[htbp]
  \begin{center}
\includegraphics[height=13cm]{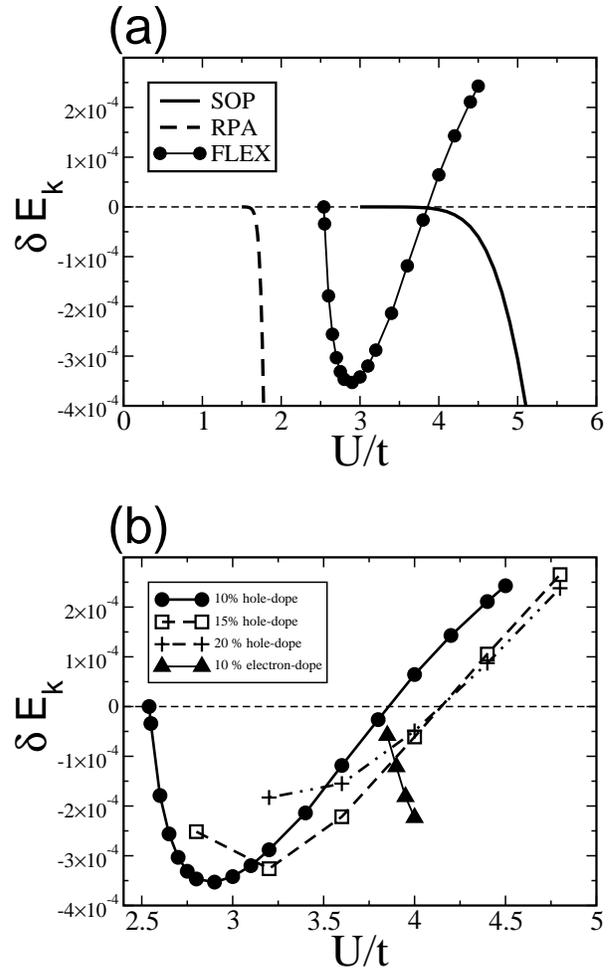}
    \caption{The difference of kinetic energy between the normal 
             state and the superconducting state. 
             (a) The results of SOP, RPA and FLEX at $\delta=0.1$. 
             (b) The results for $10$\%, $15$\% and $20$\% 
             hole-dopings as well as $10$\% electron-doping. 
             }
    \label{fig:kinetic}
  \end{center}
\end{figure}

 First, Fig.~\ref{fig:kinetic}(a) show the typical $U$-dependence 
of $\delta E_{\rm k}$ in the hole-doped region. 
 We have also performed the weak coupling theory using 
the second order perturbation theory (SOP) and 
random phase approximation (RPA). 
 While the SOP and RPA are performed at $T=0$, 
the FLEX is performed at $T=0.005$. 
 In the FLEX approximation at finite temperature, 
there is a critical value of $U$ above which 
superconductivity occurs. 
 In the present case, $U_{\rm cr}/t=2.54$ at $T=0.005$. 
 This temperature is far below \Tc if $U/t > 2.8$. 
 For example, $T_{\rm c}=0.0102$ at $U/t=4.2$.

 It is clearly shown that the sign of $\delta E_{\rm k}$ changes 
from negative to positive with increasing $U/t$. 
 The negative sign is expected in the conventional BCS theory. 
 In the BCS theory, the kinetic energy increases owing to 
the particle-hole mixing which is essential for the Cooper pairing. 
 This increase of kinetic energy is logarithmically divergent 
for the cut-off of energy as 
$\delta E_{\rm k}=-\rho \Delta^{2} \log\frac{\omega_{\rm c}}{\Delta}$, 
whose absolute value is much larger than the condensation energy  
$\delta F=\frac{1}{2} \rho \Delta^{2}$. Here, $\rho$ is the electronic 
DOS at the Fermi level. 
 In the SOP and RPA, these weak coupling behaviors are reproduced. 
 We see that the FLEX approximation also reproduces the weak 
coupling behaviors in the weak coupling region.

\begin{figure}[htbp]
  \begin{center}
\includegraphics[height=8cm]{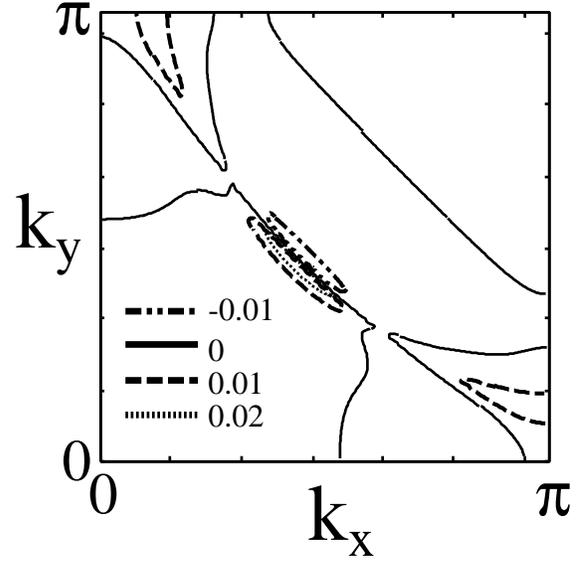}
    \caption{Contour plot of the difference of 
             momentum distribution function between 
             the superconducting state 
             and normal state, $n^{\rm S}(\k)-n^{\rm N}(\k)$. 
             We show the result for $U/t=4.2$, $\delta=0.1$ and $T=0.005$. 
             }
    \label{fig:momentum-distribution}
  \end{center}
\end{figure}

 On the other hand, the positive sign in the strong coupling region is 
a remarkably unconventional. 
 In order to clarify the microscopic origin of this behavior, 
we show the momentum distribution function 
in Fig.~\ref{fig:momentum-distribution}. 
 We see that $n^{\rm S}(\k)-n^{\rm N}(\k)$ 
takes large absolute value around the Fermi surface. 
 This is because the quasi-particles near the Fermi surface 
mainly contribute to the superconductivity. 
 The qualitatively different behavior of $\delta n(\k)$ in the 
direction perpendicular to the Fermi surface is a key to  
understand the unusual behavior. 
 It is shown that $n^{\rm S}(\k)-n^{\rm N}(\k)$ 
is positive (negative) below (above) Fermi surface at the cold spot 
around $\k=(\pi/2,\pi/2)$.  
 The situation is opposite at the hot spot around $\k=(\pi,0)$. 
 This result means that the kinetic energy arising from the hot spot 
increases owing to the particle-hole mixing induced by 
the superconducting gap.  
 On the other hand, the kinetic energy arising from the cold spot 
decreases owing to the feedback effect on the spin 
fluctuation.~\cite{rf:monthouxFLEX,rf:paoFLEX} 
 The gap in the magnetic excitation induced by the superconducting gap 
remarkably decreases the correlation effects on the low energy 
electron states. 
 Therefore, quasi-particles recover their coherent character below \Tcf. 
 Since the superconducting gap is small at the cold spot, 
this feedback effect dominates the role of particle-hole mixing. 
 The sign of $\delta E_{\rm k}$ is determined by these two competing 
effects. In the weak coupling region, the particle-hole mixing 
is dominant and $\delta E_{\rm k}$ is negative. 
 On the other hand, the feedback effect is dominant in the strong 
coupling region where $\delta E_{\rm k}$ is positive. 
 The qualitatively similar effect has been discussed phenomenologically 
as a ``quasi-particle undressing''.~\cite{rf:hirsh1992,rf:hirsh2002,
rf:norman2002,rf:haslingerrapid,rf:knigavko} 
 Here, the ``undressing'' is caused by the feedback effect on 
the spin fluctuation. 
 Note that the positive sign of $\delta E_{\rm k}$ has been reported in the 
spin-fermion model~\cite{rf:haslingerrapid,rf:haslingerfull} 
where the momentum dependence is simply neglected. 
 However, our microscopic calculation shows that 
the momentum dependence plays an essential role. 
 Note that the positive value of $\delta E_{\rm k}$ is also consistent 
with some numerical methods.~\cite{rf:yokoyamakinetic,rf:maier}

 The positive sign of $\delta E_{\rm k}$ means that the kinetic energy 
plays a role for lowering the internal energy below \Tcf. 
 This is in sharp contrast to the BCS theory, but as shown here,  
the spin-fluctuation-induced-superconductivity also gives kinetic 
energy gain. 
 It should be noticed that the FLEX approximation is a microscopic 
description of the nearly anti-ferromagnetic Fermi liquid theory, where
the Cooper pairing between quasi-particles occurs. 
 Therefore, the kinetic energy gain is not a consequence of the
non-Fermi liquid normal state as assumed in Ref.~39, and it is not a 
negative evidence for the concept of Cooper pairing between 
quasi-particles as argued in Refs.~28-30.

 Generally speaking, the kinetic energy gain can occur in the 
strong coupling superconductors where the feedback effect is important. 
 However, such a strong feedback effect as to change the sign of 
$\delta E_{\rm k}$ is not expected in the low-\Tc superconductors. 
 We point out that the gain of kinetic energy in Fig.~\ref{fig:kinetic} 
is due to (i) strong AF spin fluctuation, (ii) high-\Tcf, namely large 
superconducting gap and (iii) $d$-wave symmetry. 
 The existence of line node due to (iii) is especially important 
as is shown in Fig.~\ref{fig:momentum-distribution}.

 Next, we discuss the doping dependence. 
 Fig.~\ref{fig:kinetic}(b) shows the results of $\delta E_{\rm k}$ 
for various dopings. 
 We have shown the results in the electron-doped region as well as 
in the hole-doped region. 
 Here, the temperature is fixed to be $T=0.005$ in the hole-doped case 
and $T=0.003$ in the electron-doped case, respectively. 
 We see that the kinetic energy gain occurs in the hole-doped region 
and the crossover value of $U$ increases with increasing the doping. 
 Thus, the kinetic energy gain is likely in the under-doped region.

 It is interesting that the kinetic energy gain does not occur in the 
electron-doped region. 
 The region of superconducting state is very narrow as $U/t=3.8 \sim 4$, 
and the tendency to the superconductivity is remarkably weak 
in the electron-doped region. This is mainly because the electronic DOS 
is small and also because the spin fluctuation is sharply localized 
in the momentum space.~\cite{rf:yanaseFLEXPG,rf:yanasereview} 
 The latter indicates that the spin fluctuation is very weak. 
 Actually, an anti-ferromagnetic instability occurs at $U/t >4$ if we 
introduce a weak three-dimensionality. 
 The absence of kinetic energy gain is due to the significantly small 
magnitude of superconducting gap. 
 If we define $\Delta$ as the maximum of superconducting gap 
$\tilde{\Delta}(\k)$, we obtain nearly the BCS value 
$2 \Delta /T_{\rm c} = 4 \sim 5$ in the electron-doped region, 
while $2 \Delta /T_{\rm c} = 8 \sim 10$ in the under-doped region. 
 Combined with small value of \Tcf, the superconducting gap $\Delta$ 
in the electron-doped region is much smaller than that in the 
hole-doped region. 
 Therefore, the effect of superconducting gap on the spin fluctuation 
is not so significant. 
 We find that neither the magnetic resonance peak nor the kink in the 
electronic dispersion, which are interesting subsequences of 
feedback effect, appear in the electron-doped 
region.~\cite{rf:yanaseunpublished}

\subsection{Optical sum rule}

 The role of kinetic energy has been discussed extensively 
with the relation to the optical sum rule which is measured 
experimentally.~\cite{rf:molegraaf,rf:santander,rf:boris,rf:tajima,
rf:homes2003} 
 Here, we show that the decrease of kinetic energy can be observed by 
the measurement of optical integral, although they are 
different quantities. 

 In the isotropic system like $^{3}$He, the optical integral, 
namely the frequency integral of optical spectrum, is conserved through 
the superconducting transition as, 
\begin{eqnarray}
  \label{eq:FGT-sum-rule}
  \int_{-\infty}^{\infty} \sigma_{\rm xx}(\omega) {\rm d}\omega = 
\pi e^{2} \frac{n}{m}. 
\end{eqnarray}
 This is called Ferrell-Grover-Timkam sum rule.~\cite{rf:FGT} 
 Here, $\sigma_{\rm xx}(\omega)$ is the optical conductivity including 
the $\delta$-function at $\omega=0$, which corresponds to the 
superfluid density.  
 This sum rule is generally violated under the periodic potential, namely 
in the metals. 
 According to the Kubo formula,~\cite{rf:scalapino} 
the optical integral is related to the momentum distribution 
function $n(\k)$ as,~\cite{rf:norman2002} 
\begin{eqnarray}
  \label{eq:NP-sum-rule}
 \int_{-\infty}^{\infty} \sigma_{\rm xx}(\omega) {\rm d}\omega = 
2 \pi e^{2} \Sigma_{k} 
\frac{\partial^{2} \varepsilon(\k)}{\partial k_{\rm x}^{2}} n(\k).   
\end{eqnarray}
 If we assume $t'=0$, eq.~(\ref{eq:NP-sum-rule}) is expressed 
by the kinetic energy as, 
\begin{eqnarray}
  \label{eq:NP-sum-rule2}
\int_{-\infty}^{\infty} \sigma_{\rm xx}(\omega) {\rm d}\omega 
= -\frac{\pi e^{2}}{2} E_{\rm K}, 
\end{eqnarray} 
 This relation enables the kinetic energy to be measured experimentally.

 However, the optical integral is not expressed 
by the kinetic energy in more general case $t' \ne 0$. 
 Therefore, we define the optical energy as $E_{\rm op}=- 2 \Sigma_{k} 
(\frac{\partial^{2} \varepsilon(\k)}{\partial k_{\rm x}^{2}} 
+\frac{\partial^{2} \varepsilon(\k)}{\partial k_{\rm y}^{2}}) 
n_{\rm k}$ so that the optical integral is expressed as, 
\begin{eqnarray}
  \label{eq:NP-sum-rule3}
\int_{-\infty}^{\infty} \sigma_{\rm xx}(\omega) {\rm d}\omega 
= -\frac{\pi e^{2}}{2} E_{\rm op}. 
\end{eqnarray} 
 Note that $E_{\rm op}$ is equivalent to $E_{\rm k}$ when $t'=0$.

\begin{figure}[htbp]
  \begin{center}
\includegraphics[height=5.7cm]{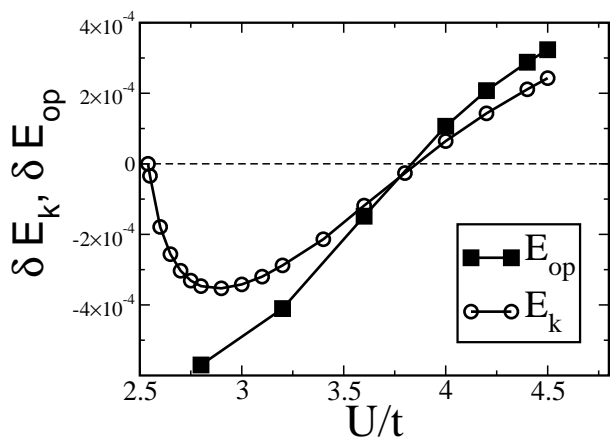}
    \caption{$U$-dependence of the optical energy $\delta E_{\rm op}$ 
             for $\delta=0.1$ and $T=0.005$. 
             We show $\delta E_{\rm k}$ for a comparison. 
             }
    \label{fig:optical}
  \end{center}
\end{figure}

 We show the $U$-dependence of $\delta E_{\rm op}$ 
in Fig.~\ref{fig:optical}. 
 It is clearly shown that the qualitative behavior of 
$\delta E_{\rm op}$ is the same as that of $\delta E_{\rm k}$ 
for a realistic value of $t'/t$. 
 The crossover of the sign occurs in common. 
 We see that the absolute value of $\delta E_{\rm op}$ is 
larger than $\delta E_{\rm k}$. 
 However, this tendency depends on the value of long 
range hoppings.~\cite{rf:comment}

 Recently, a violation of FGT sum rule has been actually observed in 
measurements of optical integral.~\cite{rf:molegraaf,rf:santander,
rf:boris,rf:tajima} 
 The results seem to be controversial, but some of them indicate 
a decrease of kinetic energy.~\cite{rf:molegraaf,rf:santander} 
 It seems that a very accurate measurement is needed since the change 
of optical integral may be much smaller than its  
absolute value.~\cite{rf:homes2003}

 Note that the maximum value of $\delta E_{\rm op}$ 
shown in Fig.~\ref{fig:optical} is $\delta E_{\rm op} \sim 0.2$meV if 
we adopt the band width $W=8 t=2$eV. 
 This value is smaller than the reported value, 
$\delta E_{\rm op} \sim 1$meV.~\cite{rf:molegraaf} 
 However, this can be a sufficient agreement because the experimental 
value significantly depends on the frequency cut-off~\cite{rf:santander} 
which is needed for a validity of the single-band description.

\subsection{Free energy and internal energy}

\begin{figure}[htbp]
  \begin{center}
\includegraphics[height=5.8cm]{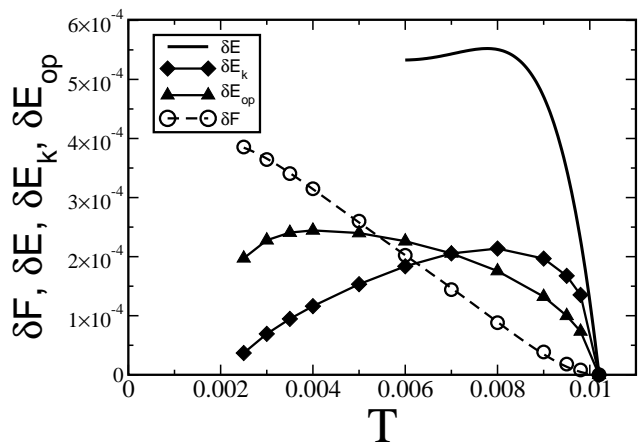}
    \caption{Temperature dependence of $\delta E$,  $\delta E_{\rm k}$, 
             $\delta E_{\rm op}$ and $\delta F$ for $\delta=0.1$ 
             and $U/t=4.2$. 
             }
    \label{fig:internal-energy}
  \end{center}
\end{figure}

 At the last of this section, we analyze the thermodynamic properties of 
spin-fluctuation-induced-superconductivity in more details. 
 First, we show the temperature dependence of internal energy 
$\delta E$, kinetic energy $\delta E_{\rm k}$, optical integral 
$\delta E_{\rm op}$ as well as the free energy $\delta F$ in 
Fig.~\ref{fig:internal-energy}. 
 We estimate the free energy by eq.~(\ref{eq:thermodynamic-super}) 
and $F=\Omega+\mu n$. 
 Since the FFT involves unphysical results around the 
cut-off frequency, we replace the summation 
of Matsubara frequency in eq.~(\ref{eq:thermodynamic-super}) as 
$\sum_{-N_{\rm f}/2}^{N_{\rm f}/2} \rightarrow 
\sum_{-N_{\rm f}/4}^{N_{\rm f}/4}$. 
 This ingenuity significantly improves an accuracy of numerical 
calculation. 
 Here, we estimate the entropy $S=-\frac{\partial F}{\partial T}$ 
by polynomial fitting and obtain the internal energy as $E=F+TS$. 
 We show the result for $\delta E$ only around $T=T_{\rm c}$ because 
the polynomial fitting is not so accurate at low temperatures.

 As is shown in the figure, we obtain the condensation energy 
$\delta F \sim 5 \times 10^{-4} \sim 0.25$meV which is consistent with 
experimental value.~\cite{rf:loram,rf:momonoprivate}  
 As for temperature dependence, the gain of kinetic energy shows 
qualitatively similar behavior to that of internal energy. 
 The decrease of $\delta E_{\rm k}$ as decreasing temperature is 
mainly owing to the decrease of kinetic energy in the normal state. 
 The damping of quasi-particle in the normal state is reduced by 
decreasing temperature, especially at the cold spot. 
 This temperature dependence of $\delta E_{\rm k}$ is qualitatively 
independent of the parameters. 
 Although $\delta E_{\rm k}$ at $T=0$ seems to be negative in Fig.~5, 
we find that $\delta E_{\rm k}$ at $T=0$ is positive 
when $U/t$ and $t'/t$ are large. 

 It is shown that a considerable part of the gain of internal energy 
is attributed to the kinetic energy, especially around $T=T_{\rm c}$. 
 However, we see $\delta E_{\rm k} < \delta E$ in the whole 
temperature region. This means that the correlation energy also 
plays a positive role for the gain of internal energy. 
 These features are robust in the intermediate coupling region 
$U/t=3 \sim 5$, but the contribution of kinetic energy increases with 
increasing $U/t$. 
 We find that most part of the condensation energy is attributed to 
the kinetic energy at $t'/t=0.35$, $U/t=7$ and $T=0$, but the kinetic 
energy gain is still smaller than the condensation energy. 
 In the variational Monte Carlo study for the Hubbard 
model,~\cite{rf:yokoyamakinetic} there is a parameter region 
$10 \leq U/t \leq 12$ where the kinetic energy and 
correlation energy play cooperative role.

 It should be noticed that the temperature dependence of optical integral 
$\delta E_{\rm op}$ is different from $\delta E_{\rm k}$, 
qualitatively. 
 Thus, it is necessary to distinguish the optical integral and kinetic 
energy when detailed properties of optical integral are discussed 
experimentally. 
 As we have shown in Fig.~\ref{fig:momentum-distribution}, 
$\delta E_{\rm k}$ and $\delta E_{\rm op}$ are determined by 
competitive contributions from the hot spot and cold spot. 
 Therefore, the detailed properties are sensitive to the long range 
hoppings.

 The specific heat over temperature is estimated from the free energy 
as $C/T=-\frac{\partial^{2} F}{\partial T^{2}}$ and 
we obtain a very large jump of specific heat as 
$(C^{\rm S}-C^{\rm N})/C^{\rm N} \sim 5.3$, 
while $(C^{\rm S}-C^{\rm N})/C^{\rm N} \sim 1$ in the weak coupling 
$d$-wave BCS theory. 
 This enhancement of specific heat jump is basically caused by 
the rapid increase of superconducting gap below \Tcf, which is 
a characteristic property of strong coupling superconductors. 
 It should be noted that the FLEX approximation is not  
appropriate in the under-doped region around $T=$\Tcf, 
because the superconducting fluctuation plays an  
important role~\cite{rf:yanasereview}. 
 Therefore, the large jump of specific heat is not observed in the 
under-doped region.~\cite{rf:loram}
 However, this large jump has been observed 
in optimally-doped region~\cite{rf:momonoprivate} as well as 
in a heavy fermion resemblance CeMIn$_{\rm 5}$.~\cite{rf:petrovic} 
where the pseudogap phenomena hardly occur.

\begin{figure}[htbp]
  \begin{center}
\includegraphics[height=5.8cm]{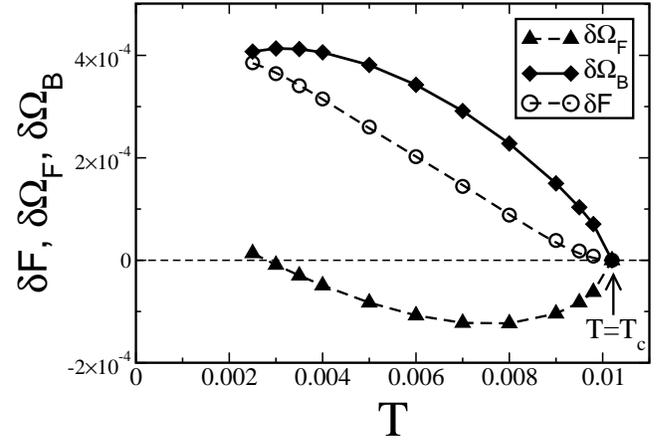}
    \caption{Temperature dependence of condensation energy $\delta F$,  
             and contributions from $\delta \Omega_{\rm F}$ and 
             $\delta \Omega_{\rm B}$. We choose $\delta=0.1$ and $U/t=4.2$. 
             }
    \label{fig:condensation-energy}
  \end{center}
\end{figure}

 Here, we propose another interpretation of the condensation energy. 
 According to eq.~(\ref{eq:thermodynamic-super}), the 
condensation energy is expressed as 
$\delta F= \delta \Omega_{\rm B}+\delta \Omega_{\rm F}+
\delta \Omega_{0}+n \delta \mu$.  
 Fig.~\ref{fig:condensation-energy} shows the contributions 
from the first and second terms. 
 The first term is expressed by eqs.~(\ref{eq:thermodynamic-superB}) 
and (\ref{eq:Phi-FLEX}). 
 If we ignore the contribution from the charge susceptibility, 
eq.~(\ref{eq:Phi-FLEX}) is equivalent to the free energy arising from 
the spin fluctuation discussed by Brinkman {\it et al}.~\cite{rf:brinkman} 
 They have estimated this term within RPA and shown that the 
Anderson-Brinkman-Morel (ABM) state in $^{3}$He is stabilized 
by the feedback effect. 
 The expression of eq.~(\ref{eq:Phi-FLEX}) is also equivalent to the 
free energy discussed in the SCR.~\cite{rf:makoshi} 
 Therefore, the first term $\delta \Omega_{\rm B}$ can be regarded 
as a free energy arising from the spin fluctuation. 
 Fig.~\ref{fig:condensation-energy} show that the first term 
$\delta \Omega_{\rm B}$  positively contributes to the condensation 
energy, while the second term $\delta \Omega_{\rm F}$ is negative 
at $T > 0.003$. 
 We see that the magnitude of $\delta \Omega_{\rm F}$ is very small 
at low temperatures and therefore the condensation energy 
is basically determined by the contribution from 
$\delta \Omega_{\rm B}$.~\cite{rf:comment2} 
 This result implies that the condensation energy mainly 
originates from the feedback effect on the spin fluctuation. 
 Note that the large spin fluctuation generally lowers $\Omega_{\rm B}$. 
 Although the static spin susceptibility $\chi_{\rm s}(q)$ at 
${\rm i}\Omega_n =0$ is reduced by the superconducting gap, 
the dynamical part at ${\rm i}\Omega_n \ne 0$ 
is enhanced by the feedback effect. 
 In the present case, the contribution from the dynamical part 
over-compensates the static part and induces the condensation 
energy of superconductivity.

 Note that this interpretation is simple, but not unique, 
since the classification of free energy in 
eq.~(\ref{eq:thermodynamic-super}) is somewhat arbitrary. 
 The spin fluctuation also affects the second term $\delta \Omega_{\rm F}$ 
through the self-energy, and this contribution is necessary so as to 
satisfy the conservation laws. 
 Nevertheless, this interpretation of condensation energy 
is somewhat interesting. 
 We see in eq.~(\ref{eq:self-energy2}) that the effective interaction 
mediated by the spin fluctuation induces the $d$-wave superconductivity. 
 Then, the static part of spin fluctuation mainly works as 
a de-pairing effect through the normal self-energy, while the 
dynamical part works as a pairing effect through the effective interaction. 
 This frequency dependence is qualitatively consistent with the above 
interpretation on the condensation energy.

 We think that this interpretation of condensation energy is 
qualitatively similar to the previous proposal~\cite{rf:scalapinoce,
rf:zhang,rf:abanovce} on the anti-ferromagnetic exchange energy 
arising from the magnetic resonance peak. 
 In the FLEX approximation, the magnetic resonance peak clearly 
appears.~\cite{rf:takimotoFLEX2} 
 Then, the frequency dependence of spin susceptibility on the real 
axis is determined by the frequency dependence on the imaginary axis. 
 The latter has been discussed above and then we found that 
the dynamical part with ${\rm i}\Omega_n \ne 0$ increases owing to 
the superconductivity. 
 This increase is an origin of magnetic resonance peak appearing in 
the spin susceptibility on the real axis. 
 Therefore, the interpretation of condensation energy discussed above 
is directly related to the appearance of the magnetic resonance peak.

\begin{figure}[htbp]
  \begin{center}
\includegraphics[height=5.8cm]{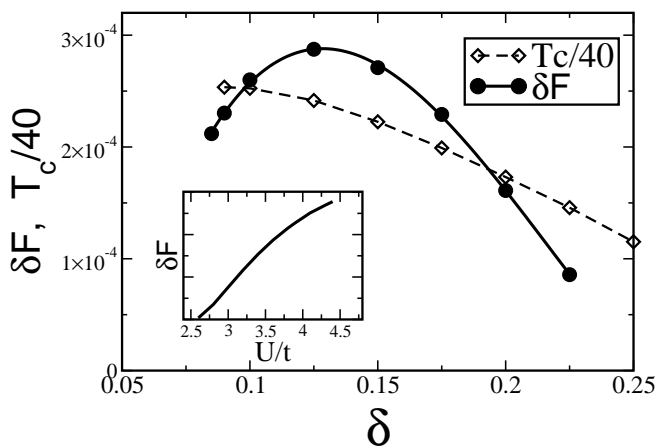}
    \caption{Doping dependence of condensation energy $\delta F$  
             for $U/t=4.2$ and $T=0.005$. 
             $T_{\rm c}/40$ is shown for a comparison.  
             The inset shows the $U$-dependence of condensation energy 
             for $\delta =0.1$ and $T=0.005$. 
             }
    \label{fig:condensation-doping}
  \end{center}
\end{figure}

 At the last of this section, we discuss the doping dependence of 
condensation energy.  
 As shown in Fig.~\ref{fig:condensation-doping}, the 
condensation energy has a dome shape. This behavior should be contrasted
to the fact that \Tc increases with under-doping. 
 The decrease of condensation energy in the under-doped region is 
basically due to the existence of competing order, namely the 
anti-ferromagnetism in the present case. In the under-doped region, 
the static spin correlation remarkably decreases the free energy 
in the normal state $F^{\rm N}$, and therefore the condensation energy 
$\delta F = F^{\rm N}-F^{\rm S}$ is reduced. 
 As shown in the inset, $\delta F$ increases monotonically 
as a function of $U$. 
 Although the phenomenological treatment has concluded that the dome shape 
appears by increasing the effective coupling constant,~\cite{rf:haslingerfull} 
our microscopic theory provides a qualitatively different result. 
 We have confirmed that these parameter dependences are consistent with 
the numerical estimation of 
eqs.~(\ref{eq:number-derivative}) and (\ref{eq:U-derivative}). 
 For example, $\mu^{\rm N}-\mu^{\rm S}$ in
eq.~(\ref{eq:number-derivative}) changes its sign around $\delta =0.125$.

 The dome shape of the condensation energy is qualitatively consistent 
with experimental observations.~\cite{rf:loram,rf:momonoprivate} 
 However, we wish to stress that our definition of condensation energy 
is somewhat different from that used in the experimental analysis. 
 Since a pseudogap exists above \Tcf, the extrapolation of the normal 
state free energy to $T <$ \Tc needs many cares. 
 If the superconducting fluctuation is an origin of pseudogap 
phenomena,~\cite{rf:yanaseFLEXPG} the free energy is 
reduced by the fluctuation above \Tcf. 
 In this case, the condensation energy may be under-estimated 
if the role of pseudogap is neglected.~\cite{rf:leggett} 
 Actually, the experiments have observed a remarkable 
suppression of condensation energy in the under-doped 
region.~\cite{rf:loram,rf:momonoprivate}

\section{Summary and Discussions}

 In this paper, we have analyzed a mechanism of high-\Tc 
superconductivity with a main interest on the energetics. 
 The Hubbard Hamiltonian was analyzed on the basis of the FLEX 
approximation. 
 It is shown that the kinetic energy is decreased below \Tc 
in the under-doped region, while it is increased in the over-doped region. 
 Interestingly, the gain of kinetic energy can not occur 
in the electron-doped region. 

 These findings are related to the violation of the optical sum rule. 
 The recent observation in the under-doped region has reported 
incompatible results to the BCS theory. However, as shown here, 
the microscopic theory on the spin-fluctuation-induced-superconductivity 
reproduces the decrease of kinetic energy in relatively strong coupling 
region, $U/t >4$. 
 The decrease of kinetic energy in the ordered state is quite unusual, 
because the phase transition from the normal state is usually induced 
by the correlation energy which competes with the kinetic energy. 
 The microscopic origin of the decrease of kinetic energy is the 
feedback effect.~\cite{rf:monthouxFLEX,rf:paoFLEX} 
 The low energy spin fluctuation significantly suppresses 
the coherent motion of quasi-particles above \Tcf, and therefore 
quasi-particle lifetime is short. Below \Tcf, the low energy spin 
fluctuation is suppressed by the opening of superconducting gap. 
 Therefore, the coherence of quasi-particles is recovered below \Tc 
especially around $(\pi/2,\pi/2)$, and the kinetic energy decreases.

 We have shown that another interpretation of condensation energy is 
possible. As discussed in \S3.3, the condensation energy is dominated by 
$\delta \Omega_{\rm B}$ which is expressed as eq.~(\ref{eq:Phi-FLEX}). 
 This term has been discussed as a contribution from the spin 
fluctuation to the condensation energy.~\cite{rf:brinkman} 
 We find that while the static part of spin fluctuation plays a 
negative role for the condensation energy, the dynamical part plays 
a positive role. 
 This interpretation is complementary with the analysis of the kinetic 
energy discussed above.

 Although we have discussed the mechanism of high-\Tc superconductivity 
from the viewpoint of the condensation energy, we wish to stress that 
the understanding obtained from the analysis of effective interaction 
is clear and useful. 
 As shown in this paper, the microscopic origin of the energy 
gain depends on the doping or $U/t$. 
 For example, $\delta E_{\rm k}$ is positive in the under-doped region, 
while it is negative in the over-doped region. 
 On the other hand, the dominant scattering process leading to 
the superconductivity is universal, namely the anti-ferromagnetic 
spin fluctuation induces the strong scattering from $(\pm \pi,0)$ to 
$(0, \pm \pi)$ which is attractive in the $d$-wave channel. 
 In other words, even if the behaviors of kinetic energy are different 
between under-doped and over-doped region, it {\it does not} mean 
that pairing mechanism changes as a function of doping.

\section*{Acknowledgments}

 The authors are grateful to H. Yokoyama for fruitful discussions and to 
N. Momono and S. Uchida for informing us of recent experimental results. 
 Numerical computation in this work was partly carried out 
at the Yukawa Institute Computer Facility. 
 The present work was partly supported by a Grant-In-Aid for Scientific 
Research from the Ministry of Education, Science, Sports and Culture, Japan.

\appendix

\section{Derivation of
eqs.~(\ref{eq:thermodynamic-super}-\ref{eq:thermodynamic-superB})}

 Here, we provide a derivation of 
eqs.~(\ref{eq:thermodynamic-super}-\ref{eq:thermodynamic-superB}) 
which are the general relation between 
the thermodynamic potential and Green function in the superconducting 
state. 
 The derivation in the normal state has been given 
by Luttinger and Ward.~\cite{rf:luttinger} 
 This can be extended to the superconducting state in a straightforward way.

 We adopt the ``Bogolyubov's trick'' in order to formulate the 
perturbation theory in the symmetry broken state. 
 We add an infinitesimal symmetry-breaking term in the Hamiltonian as 
$H \rightarrow H'$, 
\begin{eqnarray}
  \label{eq:Bogolyubov-trick}
&& \hspace{-15mm}
   H'=H'_0 + H_{\rm I},
\\
&& \hspace{-15mm}
  H'_0=\sum_{{\k},\sigma} \varepsilon(\k) 
  c_{{\k}\sigma}^{\dag}c_{{\k}\sigma}
  - \Delta_0 c_{{\k}\uparrow}^{\dag} c_{{-\k}\downarrow}^{\dag}
  - \Delta_{0}^{*}  c_{{-\k}\downarrow} c_{{\k}\uparrow}, 
\\
&&\hspace{-15mm}
  H_{\rm I} = U \sum_{i} n_{{i}\uparrow} n_{{i}\downarrow}. 
\end{eqnarray} 
 Here, $\Delta_0$ is an infinitesimal value which breaks 
the $U(1)$ gauge symmetry. We take the limit $\Delta_0 \rightarrow 0$ 
at the last of the derivation.

 We denote the Green functions and self-energy in a matrix form. 
\begin{eqnarray}
&& \hspace{-10mm} \hat{G}(k)=
 \left(
 \begin{array}{cc}
   G(k) & F(k) \\
   F^{\dag}(k) & -G(-k)
 \end{array}
 \right),
\\
&& \hspace{-10mm} \hat{\Sigma}(k)=
 \left(
 \begin{array}{cc}
   \Sigma^{\rm n}(k) & -\Delta(k) \\
   -\Delta^{*}(k) & -\Sigma^{\rm n}(-k)
 \end{array}
 \right). 
\end{eqnarray} 
 Here, the non-interacting Green function is obtained as 
\begin{eqnarray}
\hat{G}_{0}(k)=
 \left(
 \begin{array}{cc}
   G^{(0)}(k)^{-1} & \Delta_{0} \\
   \Delta_{0}^{*} & -G^{(0)}(-k)^{-1} 
 \end{array}
 \right)^{-1}. 
\end{eqnarray} 
 Then, Dyson-Gorkov equation is described as,  
\begin{eqnarray}
\hat{G}(k) = \hat{G}_{0}(k) 
                  + \hat{G}_{0}(k)  \hat{\Sigma}(k) \hat{G}(k). 
\end{eqnarray} 
 Following the perturbation theory with respect to $H_{\rm I}$, 
we obtain the differential of thermodynamic potential with respect to $U$, 
\begin{eqnarray}
\frac{d \Omega}{d U} = \frac{1}{2U} \sum_{k} 
                       {\rm Tr} \hat{G}_{0}(k) \hat{\Sigma}_{1}(k),  
\end{eqnarray} 
 where $\hat{\Sigma}_{1}(k)$ is the reducible self-energy. 
 According to the equation $\hat{\Sigma}_{1}(k) = \hat{\Sigma}(k) 
+ \hat{\Sigma}(k) \hat{G}_{0}(k) \hat{\Sigma}_{1}(k)$, we obtain, 
\begin{eqnarray}
\label{eq:differential-thermodynamic}
\frac{d \Omega}{d U} = \frac{1}{2U} \sum_{k} 
                       {\rm Tr} \hat{G}(k) \hat{\Sigma}(k).  
\end{eqnarray} 
 On the other hand, the differential of the generating function 
$\Phi[\hat{G}]$ defined by the skeleton diagram is obtained as, 
\begin{eqnarray}
\label{eq:differential-skelton}
&& \hspace{-18mm} \frac{d \Phi}{d U} = \frac{1}{2U} \sum_{k} 
                       {\rm Tr} \hat{G}(k) \hat{\Sigma}(k)
                   + \sum_{k} \frac{\delta \Phi}{\delta \hat{G}_{\mu\nu}(k)} 
                     \frac{d \hat{G}_{\mu\nu}(k)}{d U}, 
\\
\label{eq:differential-skelton2}
&& \hspace{-12mm} = \frac{1}{2U} \sum_{k} 
                       {\rm Tr} \hat{G}(k) \hat{\Sigma}(k)
                   + \sum_{k} {\rm Tr} \hat{\Sigma}(k)
                     \frac{d \hat{G}(k)}{d U}. 
\end{eqnarray} 
 Following eqs.~(\ref{eq:differential-thermodynamic}) and 
(\ref{eq:differential-skelton2}), 
\begin{eqnarray}
\label{eq:differential-differential}
&& \hspace{-20mm}\frac{d \Omega}{d U} - \frac{d \Phi}{d U} 
= -\sum_{k} {\rm Tr} \hat{\Sigma}(k)
                     \frac{d \hat{G}(k)}{d U},
\\
&& \hspace{-20mm}
= -\sum_{k} {\rm Tr} (\hat{G}_0(k)^{-1}-\hat{G}(k)^{-1})
                     \frac{d \hat{G}(k)}{d U}. 
\label{eq:differential-differential2}
\end{eqnarray} 
 Integrating the right hand side of eq.~(\ref{eq:differential-differential2})
with respect to $U$, we obtain 
\begin{eqnarray}
&& \hspace{-10mm}\Omega - \Phi - \Omega_0 = 
-\sum_{k}[\log (\frac{{\rm det}\hat{G}(k)^{-1}}
                    {{\rm det}\hat{G}_0(k)^{-1}})
+ {\rm Tr} \hat{G}(k) \hat{\Sigma}(k)]. 
\nonumber \\
&& \hspace{-10mm}
\label{eq:differential-integration}
\end{eqnarray} 
 Here, we have used the relations 
$\Omega = \Omega_0$ and $\Phi = 0$ at $U=0$. 
 Taking the limit $\Delta_0 \rightarrow 0$, we obtain  
eqs.~(\ref{eq:thermodynamic-super}-\ref{eq:thermodynamic-superB}).


\begin{thebibliography}{9}
%
\bibitem{rf:bednortz}
J. G. Bednorz and K. A. M\"uller,
Z. Phys. B {\bf64} (1986) 189.

\bibitem{rf:moriya1990} 
T. Moriya, Y. Takahashi, and K. Ueda,
J. Phys. Soc. Jpn {\bf  59} (1990) 2905; 
K. Ueda, T. Moriya, and Y. Takahashi,
J. Phys. Chem. Solids {\bf  53} (1992) 1515. 

\bibitem{rf:monthoux1991}
P. Monthoux, A. V. Balatsky and D. Pines,
Phys. Rev. Lett. {\bf  67} (1991) 3448; 
Phys. Rev. B {\bf 46} (1992) 14803.

\bibitem{rf:moriyaAD}
T. Moriya and K. Ueda, Adv. Phys. {\bf  49} (2000) 555.

\bibitem{rf:chubukovreview}
A. V. Chubukov, D. Pines, and J. Schmalian, 
{\it The Physics of Superconductors} ed. by 
K.-H. Bennemann, J. B. Ketterson, Springer, Berlin, 2002; 
A. Abanov, A. V. Chubukov, and J. Schmalian, 
Adv. Phys. {\bf 52} (2003) 119, and references therein. 

\bibitem{rf:FLEX}
N. E. Bickers, D. J. Scalapino, and S. R. White,
Phys. Rev. Lett. {\bf 62} (1989) 961;
N. E. Bickers and D. J. Scalapino,
Ann. Phys. (N.Y.) {\bf 193} (1989) 206.

\bibitem{rf:monthouxFLEX}
P. Monthoux and D. J. Scalapino, Phys. Rev. Lett. {\bf  72} (1994) 1874.

\bibitem{rf:paoFLEX}
C.-H. Pao and N. E. Bickers, Phys. Rev. Lett. {\bf  72} (1994) 1870;
Phys. Rev. B {\bf  51} (1995) 16310.

\bibitem{rf:dahmFLEX}
T. Dahm and L. Tewordt, Phys. Rev. Lett. {\bf  74} (1995) 793;
Phys. Rev. B {\bf  52} (1995) 1297.

\bibitem{rf:langerFLEX} 
M. Langer, J. Schmalian, S. Grabowski, and K. H. Bennemann,
Phys. Rev. Lett. {\bf  75} (1995) 4508.

\bibitem{rf:koikegamiFLEX} 
S. Koikegami, S. Fujimoto and K. Yamada,
J. Phys. Soc. Jpn {\bf  66} (1997) 1438.

\bibitem{rf:takimotoFLEX}
T. Takimoto and T. Moriya, J. Phys. Soc. Jpn {\bf  66} (1997) 2459.

\bibitem{rf:takimotoFLEX2}
T. Takimoto and T. Moriya, J. Phys. Soc. Jpn {\bf  67} (1998) 3570.


\bibitem{rf:yanasereview}
Y. Yanase, T. Jujo, T. Nomura, H. Ikeda, T. Hotta and K. Yamada, 
Phys. Rep. {\bf 387} (2004) 1. 

\bibitem{rf:hotta}
T. Hotta, J. Phys. Soc. Jpn. {\bf 62} (1993) 4414; 
J. Phys. Soc. Jpn. {\bf 63} (1994) 4126.

\bibitem{rf:nomura}
T. Nomura and K. Yamada, J. Phys. Soc. Jpn. {\bf 72} (2003) 2053. 

\bibitem{rf:monthoux1997}
P. Monthoux, Phys. Rev. B. {\bf 55} (1997) 15261.

\bibitem{rf:chubukov1997}
A. Chubukov, P. Monthoux and D. K. Morr,
Phys. Rev. B. {\bf 56} (1997) 7789.


\bibitem{rf:andersonbook}
P. W. Anderson, Science {\bf 235} (1987) 1196; 
P. W. Anderson,
{\it The Theory of Superconductivity in the High-$T_{\rm c}$ Cuprates}
(Princeton University Press, Princeton, 1997) and references therein.

\bibitem{rf:fukuyamaRVBreview}
H. Fukuyama, Prog. Theor. Phys. Suppl. {\bf  108} (1992) 287. 

\bibitem{rf:nagaosa} 
N. Nagaosa and P. A. Lee, Phys. Rev. Lett. {\bf 64} (1990) 2450;
Phys. Rev. B {\bf 45} (1992) 966;
P. A. Lee and N. Nagaosa, Phys. Rev. B {\bf 46} (1992) 5621.

\bibitem{rf:yokoyama}
H. Yokoyama and H. Shiba,
J. Phys. Soc. Jpn. {\bf 57} (1988) 2482; 
H. Yokoyama and M. Ogata,
J. Phys. Soc. Jpn. {\bf 65} (1996) 3615. 

\bibitem{rf:gros}
C. Gros, R. Joynt and T. M. Rice, Phys. Rev. B {\bf 36} (1987) 381. 

\bibitem{rf:dagotto}
E. Dagotto and J. Riera, Phys. Rev. B {\bf 46} (1992) 12084; 
Phys. Rev. Lett. {\bf 70} (1993) 682.

\bibitem{rf:dagottoreview}
E. Dagotto, Rev. Mod. Phys. {\bf 66} (1994) 763. 

\bibitem{rf:sorella}
S. Sorella, G. Martins, F. Becca, C. Gazza, L. Capriotti, A. Parola 
and E. Dagotto, Phys. Rev. Lett. {\bf 88} (2002) 117002. 

%
%

\bibitem{rf:hirsh1992}
J. E. Hirsh, Physica C {\bf 201} (1992) 347; 
J. E. Hirsh and F. Marsiglio, Phys. Rev. B {\bf 62} (2000) 15131. 

\bibitem{rf:imada}
M. Imada and S. Onoda, cond-mat/0008050. 

\bibitem{rf:hirsh2002}
J. E. Hirsh, Science {\bf 295} (2002) 2226. 

\bibitem{rf:chakrabarty}
S. Chakravarty, Eur. Phys. J. B {\bf 5} (1998) 337; 
S. Chakravarty, H.-Y. Kee and E. Abrahams, 
Phys. Rev. B {\bf 67} (2003) 100504. 


\bibitem{rf:molegraaf}
H. J. A. Molegraaf, C. Presura, D. van der Marel, P. H. Kes and M. Li, 
J. E. Hirsh, Science {\bf 295} (2002) 2239. 

\bibitem{rf:santander}
A. F. Santander-Syro, R. P. S. M. Lobo, N. Bontemps, Z. Konstantinovic, 
Z. Li and H. Raffy, Phys. Rev. Lett. {\bf 88} (2002) 097005; 
A. F. Santander-Syro, R. P. S. M. Lobo, N. Bontemps, Z. Konstantinovic, 
Z. Li and H. Raffy, Europhys. Lett. {\bf 62} (2003) 568. 


\bibitem{rf:anderson2000}
P. W. Anderson, Physica C {\bf 341-348} (2000) 9. 

\bibitem{rf:lee1999}
P. A. Lee, Physica C {\bf 317-318} (1999) 194. 


\bibitem{rf:maier} 
Th. A. Maier, M. Jarrell, A. Macridin and C. Slezak, 
Phys. Rev. Lett. {\bf 92} (2004) 027005. 

\bibitem{rf:yokoyamakinetic}
H. Yokoyama, Y. Tanaka, M. Ogata and H. Tsuchiura, 
J. Phys. Soc. Jpn. {\bf 73} (2004) 1119. 



%
%

\bibitem{rf:norman2000}
M. R. Norman, M. Randeria, B. Jank\'o and J. C. Campuzano, 
Phys. Rev. B {\bf 61} (2000) 14742. 

\bibitem{rf:norman2002}
M. R. Norman and C. P\'epin, 
Phys. Rev. B {\bf 66} (2002) R100506. 

\bibitem{rf:haslingerrapid}
R. Haslinger and A. V. Chubukov, 
Phys. Rev. B {\bf 67} (2003) 140504. 

\bibitem{rf:haslingerfull}
R. Haslinger and A. V. Chubukov, 
Phys. Rev. B {\bf 68} (2003) 214508. 

\bibitem{rf:eckl}
T. Eckl, W. Hanke and E. Arrigoni, 
Phys. Rev. B {\bf 68} (2003) 014505. 


\bibitem{rf:knigavko}
A. Knigavko, J. P. Carbotte and F. Marsiglio, 
Phys. Rev. B {\bf 70} (2004) 224501. 

\bibitem{rf:luttinger}
J. M. Luttinger and J. C. Ward, Phys. Rev. {\bf 118} (1960) 1417;
J. M. Luttinger, Phys. Rev. {\bf 119} (1960) 1153.

\bibitem{rf:baym} 
G. Baym and L. P. Kadanoff, Phys. Rev. {\bf  124} (1961) 287;
G. Baym and L. P. Kadanoff, Phys. Rev. {\bf  127} (1962) 1391.


%
%

\bibitem{rf:ILT}
P. W. Anderson, Science {\bf 268} (1995) 1154. 


\bibitem{rf:basov1999}
D. N. Basov, S. I. Woods, A. S. Katz, E. J. Singley, R. C. Dynes, 
M. Xu, D. G. Hinks, C. C. Homes and M. Strongin, 
Science {\bf 283} (1999) 49. 

\bibitem{rf:katz}
A. S. Katz, S. I. Woods, E. J. Singley, T. W. Li, M. Xu, D. G. Hinks, 
R. C. Dynes and D. N. Basov, Phys. Rev. B {\bf 61} (2000) 5930. 

\bibitem{rf:basov2001}
D. N. Basov, C. C. Homes, E. J. Singley, M. Strongin, T. Timusk, 
G. Blumberg and D. van der Marel, 
Phys. Rev. B {\bf 63} (2001) 134514. 


\bibitem{rf:moler}
 K. A. Moler, J. R. Kirtley, D. G. Hinks, T. W. Li, M. Xu, 
Science {\bf 279} (1998) 1193. 

\bibitem{rf:vandermarel}
A. A. Tsvetkov, D. van der Marel, K. A. Moler, J. R. Kirtley, 
J. L. de Boer, A. Meetsma, Z. F. Ren, N. Koleshnikov, D. Dulic, 
A. Damascelli, M. Gr\"uninger, J. Sch\"utzmann, J. W. van der Eb, 
H. S. Somal and J. H. Wang, Nature, {\bf 395} (1998) 360. 

\bibitem{rf:kirtley}
J. R. Kirtley, K. A. Moler, G. Villard and A. Maignan, 
Phys. Rev. Lett. {\bf 81} (1998) 2140. 

%
%
%

\bibitem{rf:manske} 
D. Manske, I. Eremin and K. H. Bennemann,
Phys. Rev. B. {\bf 62} (2000) 13922.

\bibitem{rf:yanaseFLEXPG}
Y. Yanase and K. Yamada,
J. Phys. Soc. Jpn. {\bf 70} (2001) 1659. 

\bibitem{rf:ekondo}
H. Kondo and T. Moriya, J. Phys. Chem. Solids, {\bf 63} (2002) 1399. 

\bibitem{rf:hirashima}
H. Yoshimura and D. S. Hirashima, 
J. Phys. Soc. Jpn. {\bf 73} (2004) 2057. 


\bibitem{rf:yanaseTRPG}
Y. Yanase, J. Phys. Soc. Jpn {\bf 71} (2002) 278. 

\bibitem{rf:yanasebeyond1}
Y. Yanase, J. Phys. Soc. Jpn {\bf 73} (2004) 1000. 

\bibitem{rf:yanaseSC} 
Y. Yanase, T. Jujo and K. Yamada, J. Phys. Soc. Jpn. {\bf  69} (2000) 3664. 

\bibitem{rf:yanaseunpublished} 
 We have confirmed that the kink structure in the electronic dispersion 
clearly appears in the superconducting state of under-doped region. 
 This is not the case in the electron-doped region. 

%
%

\bibitem{rf:boris}
A. V. Boris, N. N. Kovaleva, O. V. Dolgov, T. Holden, C. T. Lin, 
B. Keimer and C. Bernhard, Science {\bf 304} (2004) 708.

\bibitem{rf:tajima}
S. Tajima, Y. Fudamoto, T. Kakeshita, B. Gorshunov, V. Zelezny, 
K. M. Kojima, M. Dressel and S. Uchida, cond-mat/0401447. 


\bibitem{rf:homes2003}
C. C. Homes, S. V. Dordevic, D. A. Bonn, R. Liang and W. N. Hardy, 
Phys. Rev. B {\bf 69} (2004) 024514. 

\bibitem{rf:FGT} 
M. Tinkham, {\it Introduction to Superconductivity} 
(McGraw-Hill, 1975). 

\bibitem{rf:scalapino}
D. J. Scalapino, S. R. White and S. C. Zhang, 
Phys. Rev. B {\bf 47} (1993) 7995. 

\bibitem{rf:comment}
 We have confirmed that crossover in $\delta E_{\rm op}$ 
occurs at $t'/t=0.4$, but its absolute value is smaller than the gain of 
kinetic energy. 

\bibitem{rf:loram}
J. W. Loram, J. Luo, J. R. Cooper, W. Y. Liang, J. L. Tallon, 
J. Phys. Chem. Solids {\bf 62} (2001) 59. 

\bibitem{rf:momonoprivate}
N. Momono, T. Matsuzaki, M. Oda and M. Oda, 
J. Phys. Soc. Jpn. {\bf 71} (2002) 2832; 
T. Matsuzaki, N. Momono, M. Oda and M. Oda, 
J. Phys. Soc. Jpn. {\bf 73} (2004) 2232. 

\bibitem{rf:petrovic}
H. Hegger, C. Petrovic, E. G. Moshopoulou, M. F. Hundley, 
J. L. Sarrao, Z. Fisk, and J. D. Thompson, 
Phys. Rev. Lett. {\bf 84} (2000) 4986. 
C Petrovic, P G Pagliuso, M F Hundley, R Movshovich, J L Sarrao, 
J D Thompson, Z Fisk and P Monthoux, 
J. Phys. Condens. Matter {\bf 13} (2001) L337. 

\bibitem{rf:brinkman}
W. F. Brinkman, J. W. Seren and P. W. Anderson, 
Phys. Rev. A {\bf 10} (1974) 2386. 

\bibitem{rf:makoshi}
K. Makoshi and T. Moriya, J. Phys. Soc. Jpn. {\bf 38} (1975) 10. 


\bibitem{rf:comment2}
 The fraction of $\delta \Omega_{\rm B}$ of condensation energy depends 
on the parameters, however it is larger than 40\% 
in the under-doped region and for realistic values of $t'/t$. 

\bibitem{rf:scalapinoce}
D. J. Scalapino and S. R. White, Phys. Rev. B {\bf 58} (1998) 8222. 

\bibitem{rf:zhang}
E. Demler and S. C. Zhang, Nature {\bf 396} (1998) 733. 

\bibitem{rf:abanovce}
A. Abanov and A. V. Chubukov,  Phys. Rev. B {\bf 62} (2000) R787. 

\bibitem{rf:leggett}
D. van der Marel, A. J. Leggett, J. W. Loram and J. R. Kirtley, 
Phys. Rev. B {\bf 66} (2002) 140501. 


\end{thebibliography}
\end{document}